\documentclass[pra,twocolumn,a4paper,superscriptaddress,showpacs]{revtex4}

\usepackage{amsmath}
\usepackage{graphicx}
\usepackage{sistyle}
\usepackage{todonotes}
\usepackage{gensymb}
\usepackage{bbold}
\usepackage{bm}
\usepackage{natbib}
\usepackage{amsfonts}
\usepackage{dsfont}
\usepackage{amssymb}
\usepackage{latexsym}
\usepackage{fancyhdr}
\usepackage[bbgreekl]{mathbbol}
\usepackage{epsfig}

\def\bra#1{\mathinner{\langle{#1}|}}
\def\ket#1{\mathinner{|{#1}\rangle}}

\def\text#1{\textrm{#1}}

\DeclareMathOperator{\Tr}{Tr}

\begin{document}

%\preprint{}

\title{Revealing Genuine Optical-Path Entanglement}

\date{\today}
\author{F.~Monteiro\footnotemark[3] \footnotetext{\footnotemark[3] These authors contributed equally
to this work.}}
\affiliation{Group of Applied Physics, University of Geneva, CH-1211 Geneva 4, Switzerland}
\author{V.~Caprara Vivoli\footnotemark[3]}
\affiliation{Group of Applied Physics, University of Geneva, CH-1211 Geneva 4, Switzerland}
\author{T.~Guerreiro}
\affiliation{Group of Applied Physics, University of Geneva, CH-1211 Geneva 4, Switzerland}
\author{A.~Martin}
\affiliation{Group of Applied Physics, University of Geneva, CH-1211 Geneva 4, Switzerland}
\author{J.-D.~Bancal}
\affiliation{Centre for Quantum Technologies, National University of Singapore, 3 Science Drive 2, Singapore 117543}
\author{ H.~Zbinden}
\affiliation{Group of Applied Physics, University of Geneva, CH-1211 Geneva 4, Switzerland}
\author{R.~T.~Thew}
\email{robert.thew@unige.ch}
\affiliation{Group of Applied Physics, University of Geneva, CH-1211 Geneva 4, Switzerland}
\author{N.~Sangouard}
\email{nicolas.sangouard@unibas.ch}
\affiliation{Department of Physics, University of Basel, CH-4056 Basel, Switzerland}

\date{\today}

\begin{abstract} 
How can one detect entanglement between multiple optical paths sharing a single photon? We address this question by proposing a scalable protocol, which only uses local measurements where single photon detection is combined with small displacement operations. The resulting entanglement witness does not require post-selection, nor assumptions about the photon number in each path. Furthermore, it guarantees that entanglement lies in a subspace with at most one photon per optical path and reveals genuinely multipartite entanglement. We demonstrate its scalability and resistance to loss by performing various experiments with two and three optical paths. We anticipate applications of our results for quantum network certification.
\end{abstract}
\maketitle

Optical path entanglement -- entanglement between several optical paths sharing a single photon -- is one of the simplest forms of entanglement to produce. It is also a promising resource for long-distance quantum communication where the direct transmission of photons through an optical fiber is limited by loss. In this context, loss can be overcome by using quantum repeaters, which require the creation and storage of entanglement in small-distance links and subsequent entanglement swapping operations between the links. Among the different quantum repeater schemes, those using path-entangled states $|1\rangle_A|0\rangle_B +  |0\rangle_A|1\rangle_B$, where a single photon is delocalized into two nodes A and B are appealing - they require fewer resources and are less sensitive to memory and detector efficiencies compared to repeater architectures based e.g. on polarization entanglement~\cite{Sangouard11b}. Many ingredients composing these networks have been experimentally demonstrated, including path entanglement based teleportation~\cite{Lombardi02}, entanglement swapping~\cite{Sciarrino02}, purification~\cite{Salart10}, quantum storage~\cite{Chou05, Choi08} and an elementary network link~\cite{Chou07}. 

\begin{figure}
\includegraphics[width=0.4\textwidth]{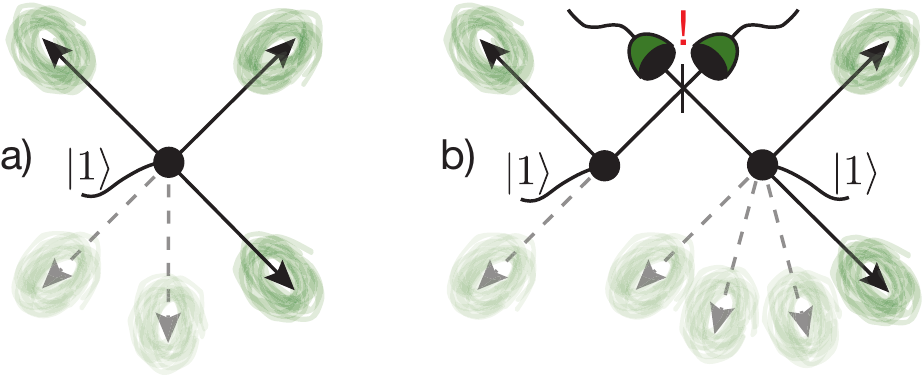}
\caption{Proposal to build up 2D networks over long distances. a) Networks made with neighboring nodes are made with $N$-path entangled states. b) These local networks can be connected remotely by means of entanglement swapping operations resulting in a large scale network.}
\label{fig:concept}
\end{figure}

A natural question is how this body of work could serve to extend known point-to-point quantum repeaters to richer geometries for quantum networks?~\figurename{~\ref{fig:concept}} presents a possible solution: A single photon incident on a multiport coupler generates entanglement over $N$ output paths (see~\figurename{~\ref{fig:concept}a}), due to its non-classical nature~\cite{Calsamiglia05}. The small network created in this way can be entangled with other, potentially distant, networks via entanglement swapping operations using 50/50 beam-splitters and single photon detectors -- a single detection is then enough to entangle the remaining 2N-2 nodes (see~\figurename{~\ref{fig:concept}b}). Such 2D networks could open up new perspectives for multi-user quantum information processing including secret sharing~\cite{Hillery99} or secure multi-party quantum computation~\cite{Crepeau02} as well as for experiments simulating quantum many-body system dynamics~\cite{Kimble08}. 

A central challenge, however, is to find an efficient, yet trustworthy, way to certify the functioning of these networks, i.e. how to characterize path entanglement in a distributed scenario using only local measurements. One might do this by using several copies of path-entangled states, as is the case for standard quantum repeater schemes~\cite{Duan01}, however, doing so is resource demanding and addresses a restrictive class of applications -- those accepting post-selection. State tomography has also been realized \cite{Babichev04} to characterize two-path entangled states but the exponential increase in measurements with the number of subsystems makes the tomographic approach impractical for detecting the entanglement in large multipartite systems like quantum networks. Recently, an entanglement witness for bipartite path entangled states has been proposed and demonstrated, that is based on a Bell inequality combined with local homodyne detections~\cite{Morin13,Ho14}. However, it is not clear how this approach can be extended to more than two paths as even for three parties we know of no Bell inequality that can be violated for W-like states with measurements lying on the equator of the Bloch sphere. \\

In this letter we propose an entanglement witness specifically developed to reveal path entanglement in distributed systems. It relies on an accurate description of measurement operators and assumes that each path is described by a single mode. However, it does not require post-selection, nor assumptions about the photon number of the measured state, hence, it reveals entanglement in a trustworthy manner. Moreover, it only makes use of local measurements and easily scales to multipartite systems. 

The principle of the witness is the following: $N$ distant observers share a state $\rho$ describing $N$ optical paths. Assuming that each path is completely described by a single mode of the electromagnetic field, the aim is not only to say whether the overall state is entangled, but also to check that entanglement lies in a subspace with at most one photon in each mode and to check that $\rho_{\bigcap\limits_{i} n_i\le1}$ is genuinely entangled. The subscript $i$ labels the observer $1, 2,.... N$ and $n_i$ is a non negative integer describing the photon number in the optical path $i$. To do this, each observer uses a measurement combining a small displacement operation and a single photon detector, a measurement initially proposed in Refs.~\cite{Tan91,Hardy94,Banaszek98} and demonstrated in~\cite{Hessmo04}. In the qubit subspace $\{|0\rangle, |1\rangle\},$ the POVM elements corresponding to click and no-click events of such a measurement can be seen as non-extremal projective measurements on the Bloch sphere whose direction depends on the amplitude and phase of the displacement \cite{Caprara14b}. In other words, if one considers non-photon-number-resolving (NPNR) detectors with a quantum efficiency $\eta$ and a small displacement $D(\alpha)=e^{\alpha a_i^\dag -\alpha^\star a_i}$  operating on the mode $i,$ the corresponding observable is given by
\begin{equation}
\bbsigma_{\alpha}^{\eta}=D^{\dagger}(\alpha)\left(2(1-\eta)^{a_i^\dag a_i}-\mathbb{1}\right)D(\alpha)\label{M}
\end{equation}
if one assigns the outcome $+1$ when the detector does not click and $-1$ when it clicks. If the measured state belongs to the subspace with at most one photon and with $\eta\,=1$, $\bbsigma_{0}$ (the superscript is omitted when $\eta=1$) corresponds to the Pauli matrix $\sigma_z$, i.e. a qubit measurement along the $\bf{z}$ direction. Similarly, for $\alpha =1$ and $\alpha=i,$ $\bbsigma_{1}$ and $\bbsigma_{i}$ are a good approximation to qubit measurements along $\bf{x}$ and $\bf{y}$, respectively. We use this analogy to build up a fidelity-based entanglement witness of the form $\mathcal{Z}_N=N(2^N|W_N\rangle \langle W_N| - \mathbb{1})$, where $W_N=\frac{1}{\sqrt{N}}\sum_{i=1}^N |0_1,..., 1_i, ...0_N\rangle$ refers to the state involving $N$ modes sharing a single photon. We approximate this expression by the operator
\begin{eqnarray}
\nonumber
&\tilde{Z}_N&=\sum_{m=1}^N (N-2m)\bbsigma_0^{\otimes m} \otimes \mathbb{1}^{\otimes N-m} \\
\nonumber
&&+  2 \sum_{m=0}^{N-2}\bbsigma_0^{\otimes m}\otimes \mathbb{1}^{\otimes N-2-m} \otimes \left(\bbsigma_{\alpha} \otimes \bbsigma_{\alpha}+\bbsigma_{i\alpha} \otimes \bbsigma_{i\alpha}\right) \\
&&+\text{sym}.
\end{eqnarray}
which only involves measurements of the form (\ref{M}). $\bbsigma_0^{\otimes m} \otimes \mathbb{1}^{\otimes N-m} $ stand for a measurement in which the first $m$ paths are measured with the observable $\bbsigma_0$ and the $N-m$ remaining ones are traced out. "sym." indicates that we add terms corresponding to permutations over all paths. 

To make our witness suitable for experiments, we focus on the case where the displacements are phase averaged so that the relative phase of displacements is random but the phase of each displacement which respect to the state on which it operates is well controlled. Under this assumption, the statistics on outcomes obtained by measuring $m$ paths with $\bbsigma_{\alpha e^{i\phi}}$ is the same for any $\phi.$ Hence, 
our witness reduces to
\begin{eqnarray}
\nonumber
&Z_N=\left(\Pi_{i=1}^{N} e^{i a_i^\dag a_i \phi}\right)& \Big(\sum_{m=1}^N (N-2m) \bbsigma_0^{\otimes m} \otimes \mathbb{1}^{\otimes N-m}   \\
\nonumber
&& + 4 \sum_{m=0}^{N-2}\bbsigma_0^{\otimes m}\otimes \mathbb{1}^{\otimes N-2-m} \otimes \bbsigma_{\alpha} \otimes \bbsigma_{\alpha} \\
&& + \text{sym}\Big) \left(\Pi_{i=1}^{N} e^{-i a_i^\dag a_i \phi}\right)\label{z_sep}
\end{eqnarray}
where $\phi$ is averaged out. In order to detect entanglement with $Z_N,$ it is suffisant to compare its value to the maximum value $z^{\max} _{\text{ppt},1} = \frac{1}{2\pi} \int_0^{2\pi} d\phi \Tr [ Z_N \rho]$ that it can take if the projection of the state $\rho$ in the $\{0,1\}$ subspace $\rho_{\bigcap\limits_i n_i\le1}$ has a positive partial transposition (PPT) with respect to a single party. Indeed, the observation of a value of $Z_N$ larger than $z^{\max} _{\text{ppt},1}$ implies by the Peres criterion \cite{Peres96,Horodecki96} that the measured state is entangled and that the entanglement lies in the qubit subspace. Since finding $z^{\max} _{\text{ppt},1}$ constitutes a linear optimization problem with semidefinite positive constraints, it can be computed efficiently (see Supplemental Material). Similarly, comparing the value of $Z_N$ to $z^{\max} _{\text{ppt}},$ the maximum value of $z^{\max} _{\text{ppt},s}$ further optimized over all possible PPTs, reveals genuine multipartite entanglement.\\

As an example, consider the value that the witness would take, $z_W$, in a scenario without loss and involving a state $W_N$ in which $N$ optical paths share a single photon. We can compare this to the value $z^{\max} _{\text{ppt}}$ that would be achieved without genuine entanglement in the $\{\ket{0}, \ket{1}\}$ subspace. We show in the Supplemental Material that
\begin{equation}
z_W-z^{\max} _{\text{ppt}}=2^{N+3}\frac{N-1}{N} |\alpha|^2e^{-2|\alpha|^2}
\end{equation}
which is positive for all $N.$ The proposed witness thus has the capability to reveal genuine entanglement of $W_N$ states for any path number. In practice, the value of $\alpha$ is optimized to make the difference $z_W-z^{\max} _{\text{ppt}}$ as large as possible. \\

When the measured state is not entirely contained in the $\{\ket{0}, \ket{1}\}$ subspace, contributions from higher photon numbers can increase the value $z^{\max} _{\text{ppt}}.$ To get a valid bound in this regime, we used autocorrelation measurements in each mode. They give a bound on the probability of having more than one photon in each path ($p_c^{(i)}$ denotes this bound for mode $i$) and avoid making assumptions about the photon number. The computation of $z^{\max} _{\text{ppt}}$ is then slightly modified to take the value of $p_c^{(i)}$ into account (see Supplemental Material). Importantly, the autocorrelation measurements are performed locally with a beam-splitter and two photon detectors.  Overall, the number of measurements required to reveal genuine entanglement between $N$ paths scales quadratically ($\frac{N^2}{2}+\frac{N}{2}+ 1$), which shows a much more favorable scaling compared to the exponential scaling of tomographic approaches.\\ 

\begin{figure}[tp]
\includegraphics[width=0.4\textwidth]{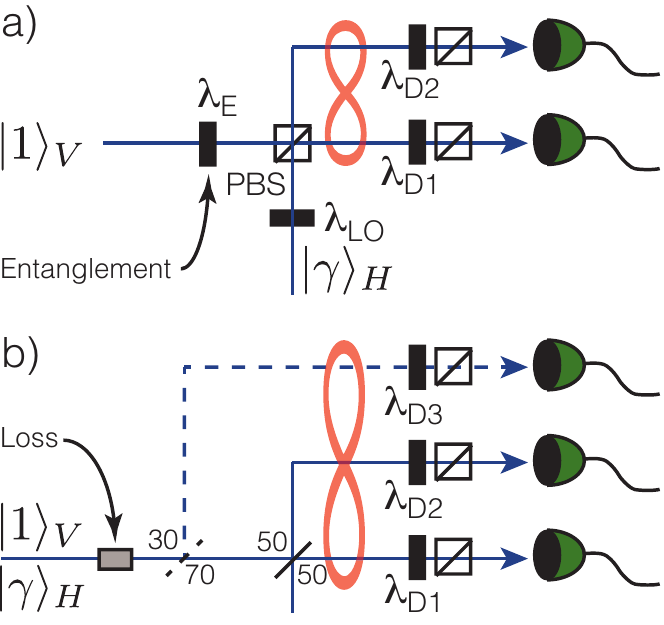}
\caption{Three different set-ups used to test the proposed entanglement witness for two and three parties: a) The heralded state can be tuned from maximally entangled to separable by the half-wave-plate (HWP) $\lambda_E$ before the first polarizing beam-splitter (PBS). The local oscillator is introduced at the other port of the PBS such that in each arm, the coherent state and the single photon have orthogonal polarization. The displacement operation is performed by rotating the HWPs at $\lambda_{D1}$ and $\lambda_{D2}$. b) The single photon and coherent state are input earlier in the set-up with orthogonal polarizations. The input loss can be varied to study the robustness of the witness. This set-up can be easily modified, by adding a 30/70 beam-splitter and another (dashed) arm, allowing us to herald and probe a tripartite W-state.}
\label{fig:schematic}
\end{figure}

We now report on a series of experiments demonstrating the feasibility of our witness. We prepare entangled networks made with 2 or 3 paths by sending single photons onto beam-splitters, see \figurename{~\ref{fig:schematic}}. The photons are prepared using a heralded single photon source (HSPS) based on a bulk PPLN nonlinear crystal~\cite{Pomarico12a}. The crystal is pumped by a pulsed laser at  532\,nm in the ps regime with a repetition rate of 430\,MHz producing non-degenerate photons at 1550\,nm and 810\,nm via spontaneous parametric down conversion. The telecom photon is filtered down to 200\,pm and subsequently detected by an InGaAs single photon avalanche diodes (SPAD), producing pure heralded single photons at 810\,nm -- the purity is verified by measuring the second order autocorrelation function $g^2(0)$~\cite{Bruno13c}. To ensure a high fidelity entangled state, the pair creation probability per pulse is limited to~10$^{-3}$, to minimize the effect of double pairs, the photons are coupled with 90\,\% efficiency~\cite{Guerreiro13b} and the overall system transmission is optimized. We herald single-photon states at a rate of $\sim$\,8\,kHz.\\

The measurements are performed by combining displacement operations and single photon detection. The local oscillator for the displacements is generated in a similar PPLN nonlinear crystal pumped by the same 532\,nm pulsed laser as well as a 1550\,nm telecom CW laser -- this ensures a high degree of indistinguishability between the HSPS and the local oscillator, which is confirmed by measuring a Hong-Ou-Mandel interference dip between the two sources, where the visibility is only limited by the statistics of the two sources~\cite{Bruno13c}. Custom gated Silicon SPADs are then used to detect the photons at 810\,nm~\cite{Lunghi12}. The detectors operate at 50\,\% efficiency and have a dark-count probability of 10$^{-2}$ per gate, for a gate width of approximately 2.3\,ns.\\

To determine the value of the witness, which reduces to 
\begin{equation}
\label{bipartite}
Z_2=\Pi_{i=1}^{2} e^{i a_i^\dag a_i \phi} \Big(2  \bbsigma_{\alpha} \otimes \bbsigma_{\alpha} -\bbsigma_{0} \otimes \bbsigma_{0} \Big) e^{-i a_i^\dag a_i\phi}
\end{equation}
in the bipartite case, we measure click/no-click (c/0) events in the two paths and calculate the corresponding probabilities, $P_{00}, P_{0c}, P_{c0}, P_{cc}$, as well as the bounds on the probabilities for having more than one photon in each path, $p_c^{(1)}$ and $p_c^{(2)}$. The correlators of the form $\{\bbsigma_{\alpha'} \otimes \bbsigma_{\alpha'}\}$ in (\ref{bipartite}) then correspond to $P_{00}+P_{cc} - P_{0c} - P_{c0}$, for $\alpha'$ = $\alpha$ or 0. We first block the single photon from going to the set-up and apply the displacement operators in both arms, validating that $|\alpha|$ corresponds to the desired value. Experimentally, $|\alpha|$ is such that $P_c \sim P_0$ locally, see Supplemental Material. Secondly we allow the single photon to go to the set-up and record the correlators with, and without, the displacement operations.  An automated series of measurements is performed, integrating over 1\,s for each setting, and is repeated as many times as needed to have good statistics. The values for  $p_c^{(1)}$ and $p_c^{(2)}$ are dominated by detector noise due to operating the detectors at such high efficiencies so as to maximize the global efficiency of the measurements. These values are used to determine the observed value of $Z_N$ labelled $z_{\rho}^\text{exp}$ and the maximum value $z^{\max}_{\text{ppt}}$ that would be obtained if the projection of the measured state in the $\{\ket{0}, \ket{1}\}$ subspace has a positive partial transpose (see Supplemental Material).

\begin{figure}[tp]
\includegraphics[width=0.4\textwidth]{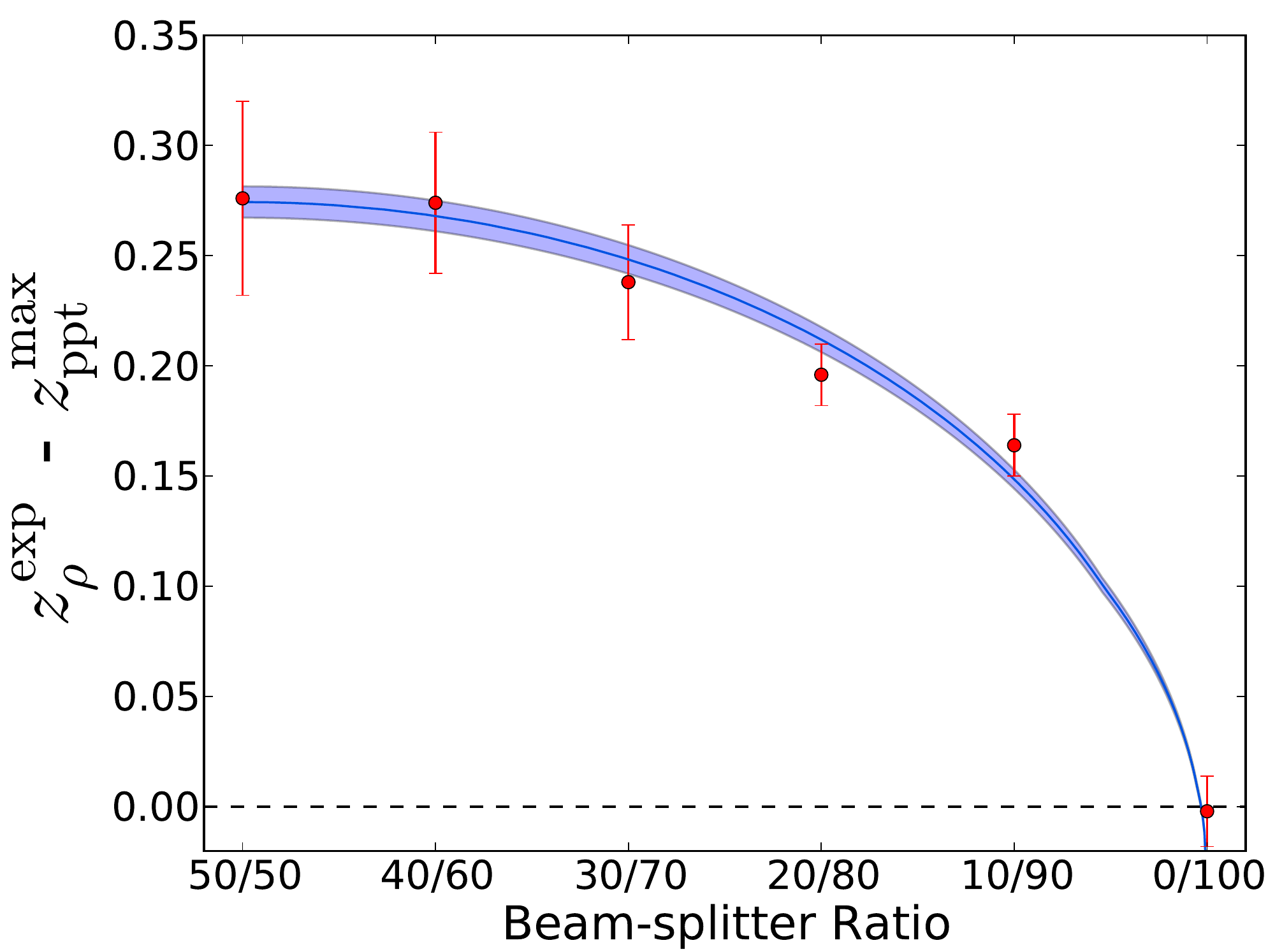}
\caption{Observed value for the bipartite entanglement witness (relatively to the PPT bound) as a function of the beamsplitter ratio. Concretely, the half waveplate $\lambda_E$ in Fig 2a) is rotated which changes the state from a maximally entangled to a separable state (50/50 - 0/100 splitting ratio, respectively). The blue band is obtained from a theoretical model, taking into account the set-up's global transmission, the characteristics of sources and gated detectors and the value for $|\alpha|\,~\sim\,0.83$ in the displacement operations.}
\label{fig:doe}
\end{figure}

To test the bipartite witness as a function of the amount of entanglement, the single photon and local oscillator are combined at different ports of a polarizing beam-splitter (PBS) ensuring that they leave in the same spatial mode with orthogonal polarizations, see \figurename{~\ref{fig:schematic}a}. A half-wave-plate (HWP) $\lambda_E$ placed in the single photon input arm is used to adjust the splitting ratio in the two output modes and the subsequent amplitudes for the entangled state.  $\bbsigma_{\alpha}$ are performed via a rotation of the wave-plates $\lambda_{D1}$, $\lambda_{D2}$ ($<$\,1 degree) before the final PBSs. The amplitude of the displacement $|\alpha|\,\sim\,0.83$ is set to maximize $z_{\rho}^\text{exp} -z_{\text{ppt}}^{\text{max}}$. \figurename{~\ref{fig:doe}} shows the result as a function of the beam-splitter ratio, from maximally entangled (50/50) to a separable state (0/100). The shaded line is obtained from a theoretical model %(including the source and detector characteristics, system transmission and detector efficiencies as well as the size of the displacement) 
with independently measured system parameters, with the associated uncertainty (see Supplemental Material). The theory and experimental results are in excellent agreement and prove that the proposed witness can reveal even very small amounts of entanglement.\\

\begin{figure}[tp]
\includegraphics[width=0.4\textwidth]{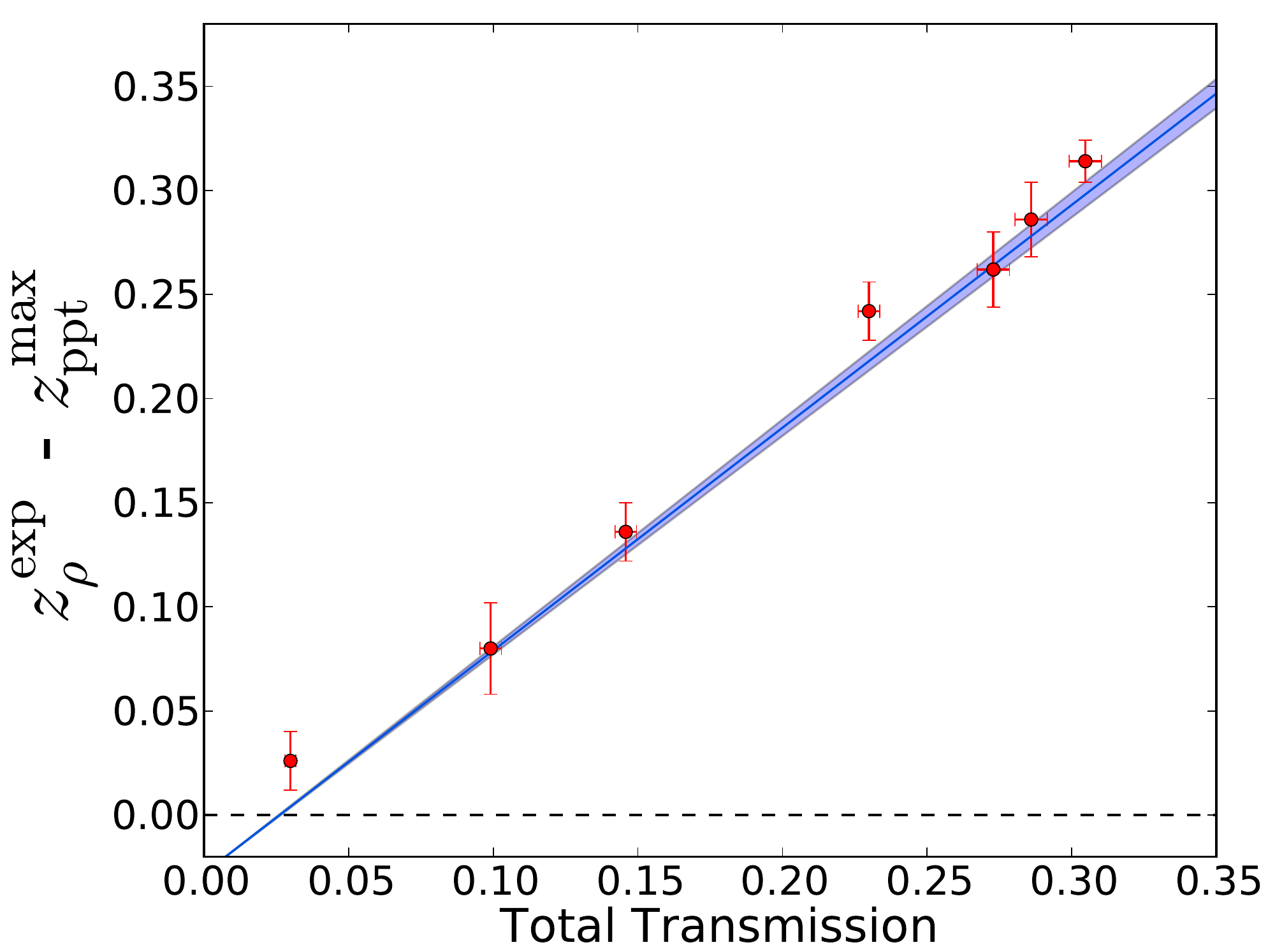}
\caption{Observed value for the bipartite entanglement witness (relatively to the PPT bound) as a function of loss. The total transmission consists of the photon coupling, transmission through the system, and detector efficiencies.}
\label{fig:dom}
\end{figure}

To prove the robustness of this witness against loss, and demonstrate the scalability, we introduce a different experimental configuration, \figurename{~\ref{fig:schematic}b}, with only a 50/50 beam-splitter, to generate maximally entangled states, and the single photon and local oscillator are combined earlier in the set-up. We can then introduce  loss to the input state, thus increasing the mixedness of the state. \figurename{~\ref{fig:dom}} shows the value of our witness of entanglement as we increase loss. The starting point has a slightly larger value than in~\figurename{~\ref{fig:doe}}, due to a slightly better transmission, n.b. the maximum transmission of $\gtrsim$30\,\% includes photon coupling, transmission and detection efficiency. Here we see that the witness is capable of revealing entanglement even in the case of high loss, or similarly, for low detection efficiency. 

Finally, by adding a 30/70 beam-splitter and another arm, dashed line in \figurename{~\ref{fig:schematic}b}, we herald tripartite states. If we assume perfect transmission and detectors with unit efficiency, we expect a maximum value for $z_{\rho}^\text{exp} -z_{\text{ppt}}^{\text{max}} \sim 7.63,$ where a value greater than zero indicates the presence of genuine entanglement. By applying our model, similarly to the bipartite case, again with $|\alpha| \sim 0.83$, but with a total transmissions in each arm of $0.19\pm0.002$, we expect to find a theoretical value of 0.99 (see Supplemental Material). We found a value of 0.99$\pm$0.10 that agrees with our model and shows a clear violation, thus revealing genuine tripartite path entanglement.\\

In conclusion, we have shown an entanglement witness suited for path entangled states that is robust and scalable, providing the means for the characterization of genuine multipartite entanglement distributed over complex quantum networks. The co-propagation of the local oscillator with the path entangled state overcomes the potential problem of distributing a phase reference, which also has the added advantage that it could be exploited for stabilisation and synchronisation of distributed networks. Interestingly, our witness provides a trustworthy means to reveal entanglement, without the need to make assumption about the number of photons in each path. A possible extension would be to make it fully device independent through the violation of a Bell inequality, which would require higher overall efficiencies~\cite{Brask13, Caprara14b}.\\

\begin{acknowledgments}
The authors would like to thank M. Ho, P. Sekatski and F. Fr\"{o}wis for stimulating discussions. This work was supported by the Swiss NCCR QSIT, the Swiss National Science Foundation SNSF (grant PP00P2-150579), the EU project SIQS and Chist-Era: DIQIP and Qscale, as well as the Singapore Ministry of Education (partly through the Academic Research Fund Tier 3 MOE2012-T3-1-009) and the Singapore National Research Foundation.
\end{acknowledgments}

%%%%%%%%%%%%%
\section*{Appendix}

\subsection{PPT bound for an arbitrary number of qubits}\label{pptbounds}
In this first paragraph, we give details on the result presented in Eq. (4) of the main text. Our witness has the following form
\begin{eqnarray}
\nonumber
&Z_N=\Pi_{i=1}^{N} e^{i a_i^\dag a_i \phi}& \Big( \sum_{m=1}^N (N-2m) \bbsigma_0^{\otimes m} \otimes \mathbb{1}^{\otimes N-m}   \\
\nonumber
&& + 4 \sum_{m=0}^{N-2}\bbsigma_0^{\otimes m}\otimes \mathbb{1}^{\otimes N-2-m} \otimes \bbsigma_{\alpha} \otimes \bbsigma_{\alpha} \\
&& + \text{sym}\Big) e^{-i a_i^\dag a_i \phi}.\label{z_general}
\end{eqnarray}
The question we address is what is the maximal value $\Tr[ \frac{1}{2\pi} \int_{0}^{2\pi} d\phi Z_N \rho]$ that $Z_N$ can take if the projection of the state $\rho$ in the $\{0,1\}$ subspace $\rho_{\bigcap\limits_i n_i\le1}$ stays positive under partial transposition with respect to some bipartition. In this section, we consider the case where each mode is described by qubits $\rho=\rho_{\bigcap\limits_i n_i\le1}$, i.e. they contain at most one photon each. This condition is relaxed in the next section. The threshold value that we look for can be obtained from the following optimization 
\begin{eqnarray}
\nonumber
&&z^{\max} _{\text{ppt},1}=\max\limits_{\rho_{\bigcap\limits_i n_i\le1}} 
 \Tr[ \frac{1}{2\pi} \int_{0}^{2\pi} d\phi Z_N \rho_{\bigcap\limits_i n_i\le1}] \-\  \text{s.t.}\\
\nonumber
&& (i) \-\  \rho_{\bigcap\limits_i n_i\le1} \ge 0,\\
\nonumber
&& (ii) \-\ \Tr \rho_{\bigcap\limits_i n_i\le1}  =1, \\
\nonumber
&& (iii) \-\ \rho_{\bigcap\limits_i n_i\le1}^{T_1}\ge 0. 
\end{eqnarray}
The two first conditions insure that $\rho_{\bigcap\limits_i n_i\le1}$ is a physical state. The last one demands that the state remains positive under partial transposition with respect to the first path. This is a linear optimization with semidefinite positive constraints which can be efficiently calculated numerically. If a physical state made with qubits leads to a larger value than $z^{\max} _{\text{ppt},1},$ we conclude that the condition (iii) does not hold, i.e. its partial transposition over the first path has at least one negative eigenvalue and is thus entangled. If we further optimize over all possible bipartitions, we obtain a bound $z^{\max} _{\text{ppt}},$ from which one can witness genuine entanglement between the $N$ optical paths. \\

\begin{figure}[!tp]
\includegraphics[width=0.40\textwidth]{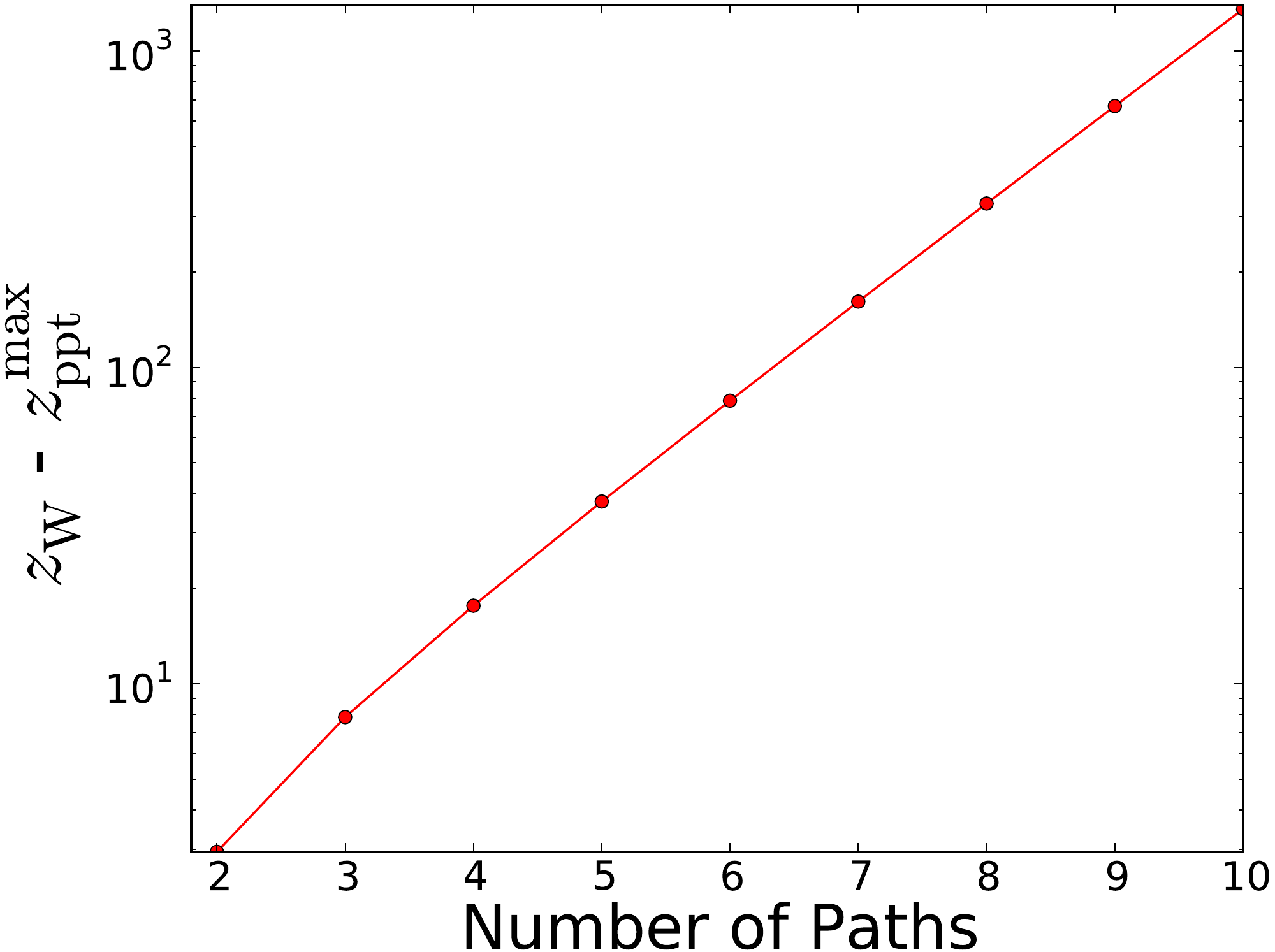}
\caption{Value that the witness takes with a $W_N$ state $z _W$ relatively to the maximal value with states that are not genuinely entangled $z^{\max} _{\text{ppt}}$ as a function of the number of optical paths. The absolute value of the difference $z _W-z^{\max} _{\text{ppt}}$ does not matter as $Z_N$ could be multiplied by any values. However, a positive difference $z _W-z^{\max} _{\text{ppt}} > 0$ reveals genuine entanglement.}
\label{fig1sm}
\end{figure}

Consider for concreteness the case where the tested state is the $W_N$ state. We can easily show that $z_W=\Tr [ \frac{1}{2\pi} \int_{0}^{2\pi} d\phi Z_N\ket{W_N}\bra{W_N}]$ is given by
\begin{eqnarray}
\label{exp_ZW}
&z_W= &(2^N-1) N + 2^{N+1} (N-1) e^{-|\alpha|^2}\\
\nonumber
&& \times \left(|\alpha|^2 (4 e^{-|\alpha|^2}-1)-1\right).
\end{eqnarray}
Furthermore, a basic measurement of the photon number distribution in each path would show that the probability of having zero photon in
 all paths or more than one photon per path is null (see below for the realization of such measurement). Together with the condition (iii), this leads to conditions on the coherence terms of the tested state, e.g. 
\begin{equation}
|\bra{10...0}\rho\ket{01...0}|^2\le p_{00...0} p_{11...0}=0.\label{Correlators}
\end{equation}
The conditions (i), (ii) (iii) together with the previously mentioned constraints allows us to get an analytical expression for the PPT bound with respect to the first path
\begin{equation}
z_W-z^{\max} _{\text{ppt},1}=2^{N+3}\frac{(N-1)}{N} |\alpha|^2e^{-2|\alpha|^2}.
\end{equation}
More generally, if we consider the entanglement between the first $m$ modes and the other $(N-m)$ modes by changing the condition over the partial transposition, we obtain
\begin{equation}
z_W-z^{\max} _{\text{ppt},m}=2^{N+3}\frac{m(N-m)}{N} |\alpha|^2e^{-2|\alpha|^2}
\end{equation}
which gives the threshold for a $m$ vs. $(N-m)$ biseparable state. This bound is identical for all such bipartitions because our witness and the conditions (i)-(ii) are invariant under exchange of parties. The threshold for genuine multipartite entanglement can be deduced by minimizing $z_W-z^{\max,m} _{\text{ppt}}$ over $m$. This leads to
\begin{equation}
z_W-z^{\max}_{\text{ppt}}=2^{N+3}\frac{N-1}{N} |\alpha|^2e^{-2|\alpha|^2}
\end{equation}
which is always positive for non-zero $|\alpha|.$ FIG.~1 is the result of the optimization of the previous expression over $\alpha$ as a function of the number of parties. This shows that our witness can detect genuine entanglement of $W_N$ states for any number of parties.\\

%%%%%%%%%%%%%

\subsection{PPT bound for qudits}
We consider the general case where no assumption is made about the number of photons in each path (no qubits). If the measured state is described by a single mode, it can be written as
\begin{equation}
\rho=\left(
\begin{array}{cc}
\rho_{\bigcap\limits_i n_i\le1}&\rho_{\text{coh}}\\
\rho_{\text{coh}}^{\dagger}&\rho_{\bigcup\limits_i n_i\ge2}\\
\end{array}
\right),
\end{equation}
where $\rho_{\bigcap\limits_i n_i\le1}$ denotes the block with at most one photon per partie as before, $\rho_{\bigcup\limits_i n_i\ge2}$ is the block in which at least one party has more than one photon and $\rho_{\text{coh}}$ denotes the coherence terms between these two blocks. Importantly, the only term in $\rho_{\text{coh}}$ leading to a non-zero contribution to our witness are those between $\rho_{\bigcap\limits_i n_i\le1}$ and $\rho_{\bigcup\limits_i n_i=2}.$ Given the maximal algebraic value $z_{\text{alg}}$ that $Z_N$ can take, we have
\begin{equation}
\begin{split}
\label{z_N_bound}
z_N\le& \Tr\left(\frac{1}{2\pi}\int_0^{2\pi}d\phi Z_N\left(\begin{array}{ccc}\rho_{\bigcap\limits_in_i\le1}&\rho_{\text{coh}}&0\\\rho_{\text{coh}}^{\dagger}&\rho_{\bigcup\limits_i n_i=2}&0\\0&0&0\end{array}\right)\right)\\
&+z_{\text{alg}}p\left(\bigcup\limits_in_i\ge 2\right)
\end{split}
\end{equation}
where $p\left(\bigcup\limits_i n_i\ge 2 \right)$ is the probability to have more than one photon in at least one path. This probability can be upper bounded using a 50/50 beam-splitter followed by two detectors in each path. Indeed, let us consider the reduced state describing, say, the first path $\rho_1.$ Its diagonal elements in the Fock basis can be written as $\sum_{n\geq 0} p_n^{(1)} |n\rangle\langle n |.$ The probability $p_c^{(1)}$ that this state leads to a two-fold coincidence after the beam-splitter is given by 
\begin{equation}
p_c^{(1)}=\sum_{n\geq 2} \frac{n}{2^n}(2^{n-1}-1) p_n^{(1)} \geq \frac{1}{2} \sum_{n\geq 2} p_n^{(1)} =  \frac{1}{2} p_{n\geq 2}^{(1)}
\end{equation}
where $p_{n\geq 2}^{(1)}$ is the probability to have strictly more than one photon in the path 1. Since $p\left(\bigcup\limits_i n_i\ge 2 \right) \leq \sum_{i=1}^N p_{n\geq 2}^{(i)},$ we have 
\begin{equation}
\label{bound_qubitspace}
p\left(\bigcup\limits_i n_i\ge 2 \right) \leq 2(p_c^{(1)}+p_c^{(2)}+...+p_c^{(N)}).
\end{equation}
The previous formula shows how to bound the probability for being outside the qubit subspace $\{\ket{0},\ket{1}\}^{\otimes N}$ with local measurements involving a 50/50 beam-splitter and two photon detectors. From Eqs. (\ref{z_N_bound}) and (\ref{bound_qubitspace}), we can deduce the PPT value in the general case of qudits from the following optimization 
 \begin{eqnarray}
 \label{z_sep}
 \nonumber
 &z^{\max} _{\text{ppt},1} \le & \max\limits_{\rho_{\bigcap\limits_i n_i\le2}} 
\Tr\left(\frac{1}{2\pi}\int_0^{2\pi}d\phi Z_N\rho_{\bigcap\limits_i n_i\le2}\right)\\
&&+2 z_{\text{alg}} (p_c^{(1)}+p_c^{(2)}+...+p_c^{(N)})
\-\  \text{s.t.}\\
\nonumber
& (i) & \-\  \rho_{\bigcap\limits_i n_i\le2} \ge 0,\\
\nonumber
& (ii) & \-\ \Tr \rho_{\bigcap\limits_i n_i\le2}  \leq 1, \\
\nonumber
& (iii) & \-\ \rho_{\bigcap\limits_i n_i\le1}^{T_1}\ge 0. 
\end{eqnarray}
Note first that the optimization is performed over the set of physical states (condition (i)) having two photons at maximum in each path because the only coherence terms having a non-zero contribution lie in this subspace. To keep the optimization general, the trace of the state is required to be smaller than, or equal to, one (condition (ii)). The last condition ensures that the projection of the state with at most one photon in each path has a positive partial transpose when the transposition is taken over the first path. By changing the PPT condition (iii) to $\rho_{\bigcap\limits_i n_i\le1}^{T_m} \geq 0$ where $T_m$ denotes partial transposition with respect to the bipartition $m,$ we get a PPT bound for this bipartition $z^{\max} _{\text{ppt},m}.$ The value of the PPT bound we are interested in is $z^{\max}_{\text{ppt}} = \max_m z^{\max} _{\text{ppt},m}.$ Importantly, the difference between $z_W$ and $z^{\max}_{\text{ppt}}$ is made larger by using the results of the joint probability to have clicks/no-clicks without displacement and adding the corresponding constraints in the optimization procedure, cf below. Overall, the number of required measurements $(1+C_2^N)+N=N^2/2+N/2+1$ has a quadratic scaling in the number of paths and is thus suited to prove genuine entanglement in multiple-path states. \\

%%%%%%%%%%%%%

\subsection{Witnessing path entanglement with non-unit efficiency detectors}
Note that so far we have assumed that the photon detectors have unit efficiencies. How can we use this witness in practice, when non-unit efficiency detectors are used? First, note that a detector with efficiency $\eta$ can be seen as a unit-efficiency detector preceded by a beam-splitter with a transmission efficiency  $\eta.$ Let $U_{\text{bs}}$ be the unitary corresponding to this beam-splitter. Let also $D(\alpha)$ be the unitary associated to the displacement operation with argument $\alpha$. We have 
\begin{equation}
D(\alpha \sqrt{\eta}) U_{\text{bs}} = U_{\text{bs}} D(\alpha) 
\end{equation}
meaning that the inefficiency of the detector can be modeled as a beam-splitter operating before the displacement operation provided that the amplitude of the displacement is reduced by $\sqrt{\eta}.$ Let us consider the configuration where the loss operates before the displacement. The optimizations performed so far allows one to conclude that the state after the loss is entangled. This implies that the state was already entangled before the loss as a separable state remains separable under loss. Our witness thus proves entanglement when non-unit efficiency detectors are used provided that the displacement accounts for the reduced detection efficiency. \\

%%%%%%%%%%%%%

\subsection{Bipartite case}
We now give an explicit expression of the PPT bound for the bipartite case. For $N=2,$ we take the following witness 
\begin{equation}
Z_2=\Pi_{i=1}^{2} e^{i a_i^\dag a_i \phi} \Big(2  \bbsigma_{\alpha} \otimes \bbsigma_{\alpha} -\bbsigma_{0} \otimes \bbsigma_{0} \Big) e^{-i a_i^\dag a_i\phi}.
\end{equation}
Note that we have removed a factor 2 from the definition (\ref{z_general}). 
To get the PPT bound $z^{\text{max}}_{\text{ppt}},$ first note that the maximal (algebraic) value for $Z_2$ is $3$ and since $p\left(n_1\ge2\cup n_1\ge2\right)\le 2(p_c^{(1)}+p_c^{(2)})$, we have\\
 \begin{eqnarray}
 \nonumber
 &z^{\text{max}}_{\text{ppt}} & \le\max\limits_{\rho_{n_1\le2\cap n_2\le2}}\Tr\left(\frac{1}{2\pi}\int_0^{2\pi} d\phi Z_2\rho_{n_1\le2\cap n_2\le2}\right)\\
 \nonumber
 &&+6(p_c^{(1)}+p_c^{(2)})
 \end{eqnarray}
 s.t.
\begin{enumerate}
\item $\rho_{n_1\le2\bigcap n_2\le2}\ge 0$,
\item $\Tr(\rho_{n_1\le2\bigcap n_2\le2})\le 1$,
\item $\rho_{n_1\le1\bigcap n_2\le1}^{T_2}\ge 0,$
\item $1-p \left(n_1\ge 2\bigcup n_2\ge 2 \right) = \Tr(\rho_{n_1\le1\bigcap n_2\le1}).$
\end{enumerate}
 When $\bbsigma_{0} \otimes \bbsigma_{0}$ is measured, one accesses the joint probabilities of having click/no-click without displacement. Let $P_{ij}$ be the joint probability to have outcomes $i$ and $j$ respectively, $i,j =0$ for no-click and $i,j =c$ for click. The $P_{ij}$s provide upper bounds of the diagonal terms of the measured state
$p_{00}=P_{00}$, $p_{01}\le P_{0c}$, $p_{10}\le P_{c0}$ and $p_{11}\le P_{cc}.$ Similarly, from the local probabilities of having more than one photon, we have $p_{02}\le 2 p_c^{(2)},$ $p_{20}\le 2 p_c^{(1)},$ $p_{12}\le 2 p_c^{(2)},$ $p_{21}\le 2 p_c^{(1)}$ and $p_{22}\le 2 p_c^{(1)}.$
Together with the condition $1.$ and $3.$ (implying e.g. $|\langle 10 |\rho_{n_1\le2\bigcap n_2\le2} |01\rangle |^2 \le p_{01} p_{10}$ and $|\langle 10 |\rho_{n_1\le2\bigcap n_2\le2} |01\rangle |^2 \le p_{00} p_{11}$ respectively), these constraints provide the following upper bound on $z^{\text{max}}_{\text{ppt}}$ in the regime $\alpha\ge0.45$ 
\begin{equation}
\label{z_sep2}
\begin{split}
&z^{\text{max}}_{\text{ppt}}\le \left(2\left(-1+2e^{-|\alpha|^2}\right)^2-1\right)P_{00}\\
&+\left(2\left(-1+2e^{-|\alpha|^2}\right)\left(-1+2e^{-|\alpha|^2} |\alpha|^2\right)+1\right)(P_{0c}+P_{c0})\\
&+ 2\left(2 \left(-1 + e^{-|\alpha|^2} |\alpha|^4\right)^2 - 1\right)p_c^{(1)}\\
&+2\left(2 \left(-1 + 2 e^{-|\alpha|^2}\right) \left(-1 + e^{-|\alpha|^2} |\alpha|^4\right) + 4\right)(p_c^{(1)} + p_c^{(2)})\\
&+ 16 |\alpha|^2e^{-2|\alpha|^2}\Big(\sqrt{p_c^{(1)} p_c^{(2)}} |\alpha|^4 +\sqrt{P_{00} P_{cc}}  \\
&+\left(\sqrt{p_c^{(2)} P_{cc}} + \sqrt{p_c^{(1)} P_{cc}}+ \sqrt{p_c^{(1)} p_c^{(2)}}\right) |\alpha|^2  \Big).
\end{split}
\end{equation}
This analytical bound is used and compare to the observed value of our witness. If the latter is larger than the former, we can safely conclude that the tested state is entangled.

%%%%%%%%%%%%%

\subsection{Theoretical model}
The aim of this section is to detail the model that has been used to reproduce the experimental results (blue lines in FIG. 3 and 4 of the main text). The starting point is the pair production in two different modes by means of a spontaneous down conversion source and subsequent photon detection of one of two modes. The state conditioned to a click on the heralding detector is given by \cite{Caprara14b}
\begin{equation}\begin{split}
\rho_h&=\frac{1-R_h^2T^2_{g}}{T^2_{g}\left(1-R_h^2\right)}\Big[\rho_{\text{th}}\left(\bar{n}=\frac{T_g^2}{1-T_g^2}\right)\\
&-\frac{1-T_g^2}{1-R_h^2T_g^2}\rho_{\text{th}}\left(\bar{n}=\frac{R_h^2T_g^2}{1-R_h^2T_g^2}\right)\Big],\label{EquThermalState}
\end{split}\end{equation}
i.e. a difference between two thermal states $\rho_{\text{th}}(\bar{n})$ with $\bar{n}$ mean photons. $T_g=\tanh(g)$, $g$ being the squeezing parameter, $a^{\dagger}$ and $a$ stand for bosonic operators and $R_h=\sqrt{1-\eta_h}$ where $\eta_h$ is the heralding efficiency. Importantly, a thermal state can be written as a mixture of coherent states $\ket{\gamma}.$ Concretely, $\rho_{\text{th}}\left(\bar{n}\right)=\int d^2 \gamma P^{\bar{n}}(\gamma)\ket{\gamma}\bra{\gamma}$
with $P^{\bar{n}}(\gamma)=\frac{1}{\pi \bar{n}}e^{-\frac{|\gamma|^2}{\bar{n}}}$. The correlators that we want to calculate can thus be obtained from the behavior of a coherent state. A beam-splitter splits a coherent state into two coherent states, i.e. $\ket{\gamma}\rightarrow\ket{\sqrt{R}\gamma}_a\ket{\sqrt{T}\gamma}_b$, where $T$ and $R$ are, respectively, the transmittivity and reflectivity. A displacement $D(\alpha)$ on a coherent state $\ket{\gamma}$ gives another coherent state with mean photon number $|\gamma+\alpha|^2$. Together with 
\begin{equation}
x^{\frac{a^{\dagger}a}{2}}\ket{\bar{\gamma}}=e^{-\frac{(1-x)|\bar{\gamma}|^2}{2}}\ket{\sqrt{x}\bar{\gamma}},
\end{equation}
we easily obtain the probability to get no click in both sides from a thermal state $\rho_{\text{th}}(\bar{n})$ knowing the amplitude of the local displacement $\alpha$
\begin{equation}
P_{00}^{\alpha}=\frac{e^{-2|\alpha|^2+\frac{\bar{n}\eta_t\eta}{1+\bar{n}\eta_t\eta}(\sqrt{R}+\sqrt{T})^2|\alpha|^2}}{1+ \bar{n}\eta_t\eta}
\end{equation}
where $\eta_t\eta$ is the overall efficiency (from the source to the detector, including the detection efficiency).
Following the same line of thought for $P_{cc}^{\alpha}, P_{0c}^{\alpha}, P_{c0}^{\alpha},$ we find 
\begin{equation}
\nonumber
\begin{split}
\Tr(\rho_{\text{th}}(\bar{n}) \bbsigma_{\alpha} \otimes \bbsigma_{\alpha} )
=&1+4\frac{e^{-2|\alpha|^2+\frac{\bar{n}\eta_t\eta}{1+\bar{n}\eta_t\eta}(\sqrt{R}+\sqrt{T})^2|\alpha|^2}}{1+\bar{n} \eta_t\eta}\\
-&2\frac{e^{-\frac{|\alpha|^2}{1+\eta_t\eta \bar{n} R}}}{1+\eta_t\eta\bar{n}R}-2\frac{e^{-\frac{|\alpha|^2}{1+\eta_t \eta \bar{n} T}}}{1+\eta_t \eta \bar{n}T}.
\end{split}\end{equation}
From this last expression, we deduce the correlator for the state (\ref{EquThermalState})
\begin{eqnarray}
&\Tr&(\rho_h \bbsigma_{\alpha} \otimes \bbsigma_{\alpha})=\frac{1-R_h^2T^2_{g}}{T^2_{g}\left(1-R_h^2\right)} \times \\
\nonumber
&&\Big[\Tr\left(\rho_{\text{th}}\left(\frac{T_g^2}{1-T_g^2}\right) \bbsigma_{\alpha} \otimes \bbsigma_{\alpha} \right)\\
\nonumber
&&-\frac{1-T_g^2}{1-R_h^2T_g^2}\Tr\left(\rho_{\text{th}}\left(\frac{R_h^2T_g^2}{1-R_h^2T_g^2}\right) \bbsigma_{\alpha} \otimes \bbsigma_{\alpha}\right)\Big].\label{EquThermalState2}
\end{eqnarray}
Finally, the value taken by our entanglement witness is obtained from $2\Tr(\rho_h \bbsigma_{\alpha} \otimes \bbsigma_{\alpha})-\Tr(\rho_h \bbsigma_{0} \otimes \bbsigma_{0}).$ An independent characterization of the source together with the measurements of efficiencies $\eta_h$ and $\eta_t\eta$ shows a very good agreement between our model and the results of the experiments (see FIG.~3 and 4 of the main text). The error associated to both the theoretical models in FIG.~3 and 4 of the main text take into account the error on the value of $\alpha$ plus the error in the transmission of the set-up for FIG.~3 and in the beam-splitter ratio that prepares the state for FIG.~4.\\

%%%%%%%%%%%%%

\subsection{Tripartite case}
For $N=3,$ the expression of the witness $Z_{3}$ is
\begin{eqnarray}
\nonumber
&Z_{3}=&\Pi_{i=1}^{3} e^{i a_i^\dag a_i \phi} \Big(\bbsigma_0 \otimes \left(\mathbb{1} \otimes \mathbb{1}-\mathbb{1} \otimes \bbsigma_0-3\bbsigma_0\otimes \bbsigma_0\right)\\
&&+4\left(\mathbb{1}+\bbsigma_0\right)\otimes \bbsigma_{\alpha} \otimes \bbsigma_{\alpha}+\text{sym} \Big)e^{i a_i^\dag a_i \phi}.
\label{EquZ3}
\end{eqnarray}
where sym stands for the terms obtained by all possible permutations of modes 1, 2, and 3. Following the same line of thought as for the bipartite case, we can show that for any $\alpha \ge 0.67$ 

\begin{figure}[t!]
\includegraphics[width=0.44\textwidth]{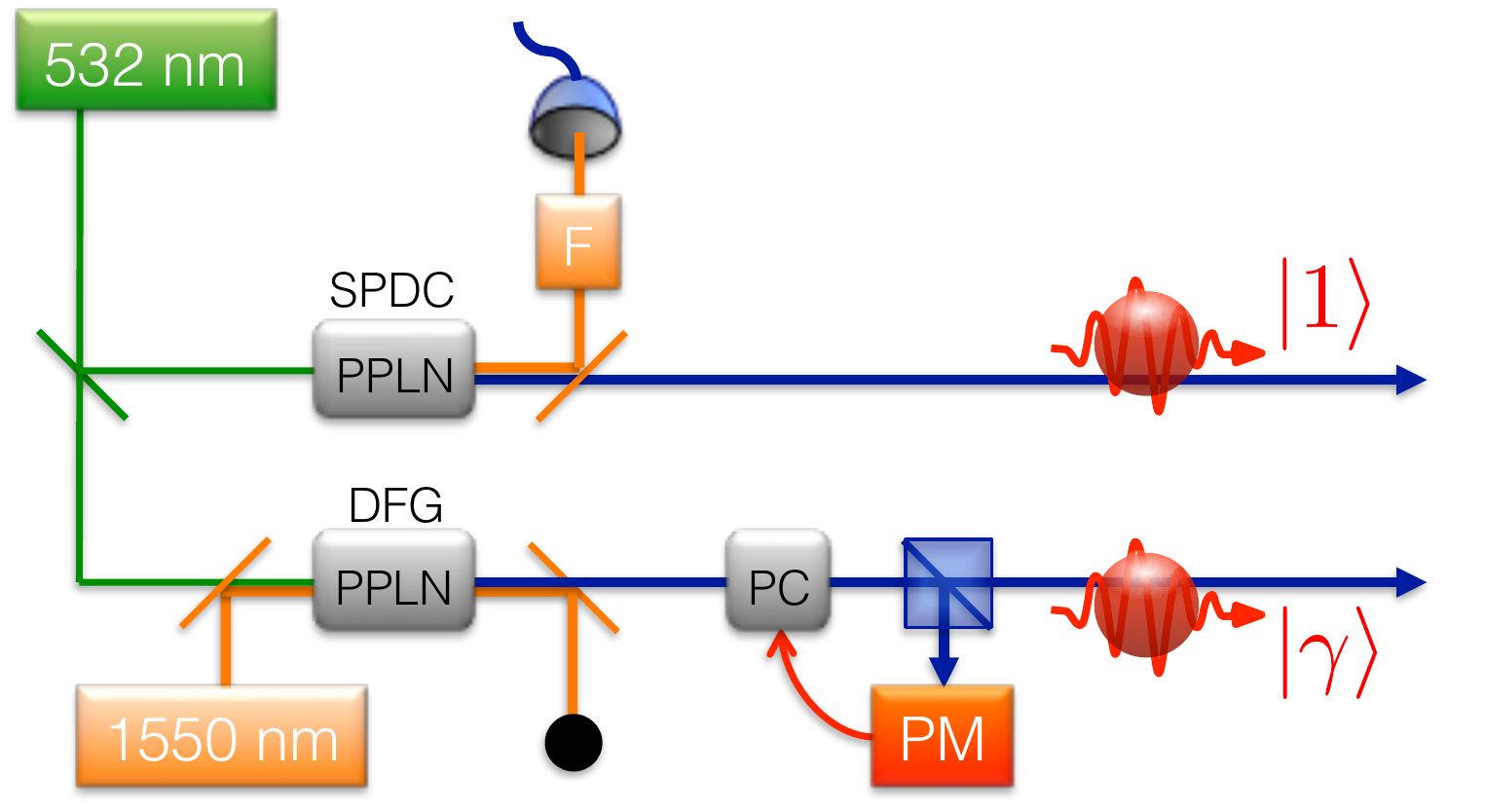}
\caption{The experimental set-up for the heralded single photon source and the local oscillator. See text for details.}
\label{fig2sm}
\end{figure}

\begin{widetext}
\begin{align}
\nonumber
z^{\text{max}}_{\text{ppt}}& \le \left(-3+24\left(-1+2e^{-|\alpha|^2}\right)^2\right) P_{\text{000}}+\left(5+16\left(-1+2e^{-|\alpha|^2}\right)\left(-1+2e^{-|\alpha|^2} |\alpha|^2\right)\right)\times (P_{\text{00c}} + P_{\text{0c0}} + P_{\text{c00}} )\\
\nonumber
&+4\left(1+4 \left(-1 + e^{-|\alpha|^2} |\alpha|^4\right) \left(-3 + 4 e^{-|\alpha|^2} +e^{-|\alpha|^2} |\alpha|^4\right)\right)\times\left(p_c^{(1)}+p_c^{(2)}+p_c^{(3)}\right) \\
\nonumber
&+64 |\alpha|^2e^{-2 |\alpha|^2}\Big(|\alpha|^2
\Big(\sqrt{p_c^{(3)} P_{\text{0cc}}}+\sqrt{p_c^{(3)} P_{\text{c0c}}}
+\sqrt{p_c^{(2)}P_{\text{0cc}}} +\sqrt{p_c^{(2)} P_{\text{cc0}}} +\sqrt{p_c^{(1)}P_{\text{c0c}}}+\sqrt{p_c^{(1)}P_{\text{cc0}}}\Big)\\
\nonumber
&\phantom{64 |\alpha|^2e^{-2 |\alpha|^2}\Big(} + |\alpha|^2\left(1 + |\alpha|^2\right) \left(\sqrt{p_c^{(3)} p_c^{(2)}} +\sqrt{p_c^{(3)} p_c^{(1)}} + \sqrt{p_c^{(2)} p_c^{(1)}}\right)\\
\nonumber
&\phantom{64 |\alpha|^2e^{-2 |\alpha|^2}\Big(} + \text{max}\Big(\sqrt{P_{\text{0c0}} P_{\text{00c}}} + \sqrt{P_{\text{000}} P_{\text{c0c}}} + \sqrt{P_{\text{000}} P_{\text{cc0}}}, \sqrt{P_{\text{c00}} P_{\text{00c}}} + \sqrt{P_{\text{000}} P_{\text{0cc}}} + \sqrt{P_{\text{000}} P_{\text{cc0}}},\\
\nonumber
 &\phantom{64 |\alpha|^2e^{-2 |\alpha|^2}\Big(+\max} \sqrt{P_{\text{0c0}} P_{\text{c00}}} + \sqrt{P_{\text{000}} P_{\text{c0c}}} + \sqrt{P_{\text{000}} P_{\text{0cc}}}\Big)\Big)\\
 \nonumber
&+33\Big(2\left(p_c^{(1)}+p_c^{(2)}+p_c^{(3)}\right)-\text{max}\left(0,2\left(p_c^{(1)}+p_c^{(2)}\right)-1,2\left(p_c^{(1)}+p_c^{(3)}\right)-1,2\left(p_c^{(2)}+p_c^{(3)}\right)-1\right)\Big)
\end{align}
\end{widetext}
Observation of a value larger than $z^{\text{max}}_{\text{ppt}}$ allows one to conclude that the measured state is genuinely entangled.

\subsection{Experiment}

FIG.~2 shows the experimental scheme for the generation of heralded single photons and the local oscillator for the displacement operations. Most of the experimental parameters are explained in the main text. We have two independent sources based on PPLN non-linear crystals, with type-0 quasi-phasematching. The heralded single photon source (HSPS) uses a tunable filter (F) with a 200\,pm bandwidth that has around 50\% transmission for the central (telecom) wavelength. This is sufficiently narrow so as to herald the 810\,nm photons in pure states~\cite{Bruno13c}. By stimulating the DFG process in the second non-linear crystal with a coherent CW laser at the same wavelength of the HSPS's filter, energy conservation dictates that both the heralded single photon and the local oscillator (LO) state are indistinguishable.  A delay arm (not shown) allows for the sychronization of the two sources. \\

To determine the value of $|\alpha|$ which maximizes $z_{\rho}^\text{exp}-z_{\text{max}}^{\text{ppt}},$ the probabilities $P_{00}, P_{0c}, P_{c0}, P_{cc}$, as well as $p_c^{(1)}$ and $p_c^{(2)}$ are measured. By guessing that they are obtained from a $W$ state, these probabilities allow one to estimate the overall efficiencies and hence the value that the witness would take for any $\alpha$. Maximizing the difference with (\ref{z_sep2}) results in $|\alpha|=0.83.$ We emphasize that in practice, with non-unit efficiency detectors, this corresponds to a displacement with amplitude $0.83/\sqrt{\eta}.$ This means that the detection efficiency does not need to be known. The values of $|\alpha|$ are locally set to produce click/no-click with the same probability independently of the detection efficiency.\\

\begin{figure}[!h]
\includegraphics[width=0.40\textwidth]{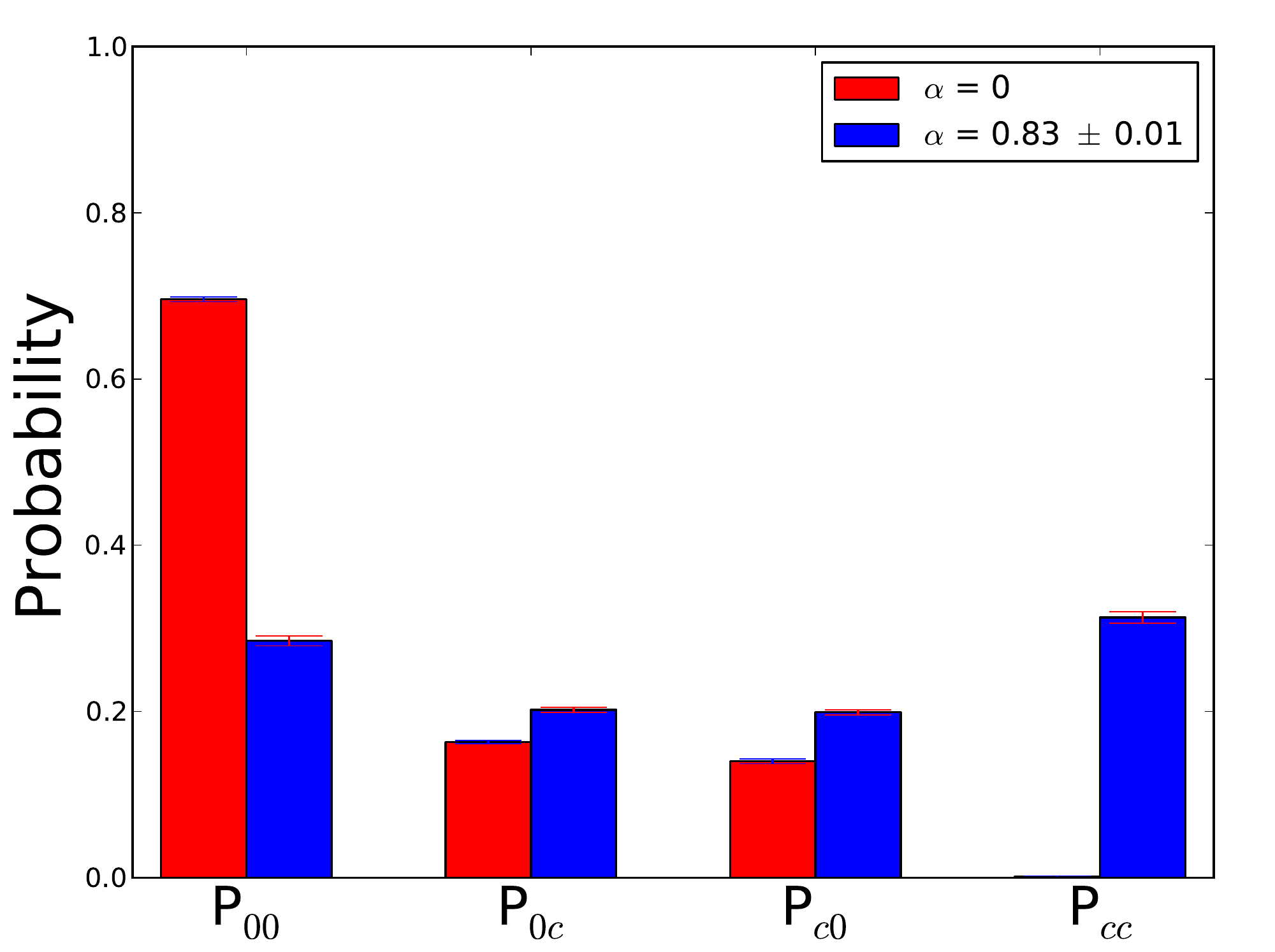}
\caption{Measured probabilities of having click/no-click in the two arms of the bipartite experiment (FIG. 4 of the main text with the maximal transmission efficiency $(\geq 0.3)$). Red columns give the probabilities measured without the displacement operation and the blue columns correspond to the probabilities measured when the displacement operator was applied ($|\alpha|=0.83\pm0.01$).}
\label{fig3sm}
\end{figure}

The stability of the mean photon number for the LO states is realized by a feedback loop, where a large fraction of the generated LO is sent, via a polarizing beam-splitter (PBS) to a power meter (PM) that then acts on a Pockels cell. The typical integration times, per point in FIG.\,3 \& 4 in the main text, are around 1.5 hours. The error bars are not Poissonian as they are dominated by fluctuations in the system, in particular, the generation of the LO, which depend on the feedback loop.

If we consider the measurement for the maximally entangled bipartite case as an example, the probability to have more than one photon in each arm are measured to be $p_c^{(1)}=10^{-4}$ and $p_c^{(2)}=10^{-4}$. After measuring the different click/no-click events we can determine their probabilities - see the results presented in FIG.~\ref{fig3sm}. These lead to $z_{\rho}^\text{exp} = -0.002$, $z^{\max}_{\text{ppt}}=-0.315,$ i.e. $z_{\rho}^\text{exp}-z^{\max}_{\text{ppt}}=0.313\,>\,0$. As the observed value was larger than $z_{\text{max}}^{\text{ppt}},$ we can conclude that the measured state is genuinely entangled.

\bibliography{Witness}

\end{document}